\documentclass[pra,twocolumn,amsmath,amssymb,floatfix,reprint,footinbib,superscriptaddress,longbibliography,showkeys]{revtex4-2}
\usepackage{picinpar,graphicx}
\usepackage{braket}
\usepackage{amsmath}
\usepackage{graphicx}
\usepackage{epstopdf}
\usepackage{dcolumn}
\usepackage{graphicx}
\usepackage{mathrsfs}
\usepackage{mdwlist}
\usepackage{subfigure}
\usepackage{booktabs}
\usepackage{amsmath}
\usepackage{dsfont}
\usepackage{amstext}
\usepackage{amssymb}
\usepackage{amsbsy}
\usepackage{bbm}
\usepackage{amsthm}
\usepackage{graphicx}
\usepackage{textcomp}
\usepackage{multirow}
\usepackage{color}
\usepackage[colorlinks,citecolor=blue]{hyperref}
\definecolor{Dgreen}{RGB}{0, 100, 0}
\usepackage{url}
\usepackage[colorlinks]{hyperref}
\hypersetup{%
	plainpages=true,
	breaklinks=true,  
	hypertexnames=false,  
	pageanchor=true,
	colorlinks=true,
	linkcolor={blue},
	citecolor={red},
	urlcolor={blue},
	anchorcolor={black}
}

\begin{document}

	\title{Entanglement generation of arbitrary squeezed Fock states}

    \author{Qin-Ru Cheng}
    \thanks{These authors contributed equally to this work.}
	\affiliation{Department of Physics, Fuzhou University, Fuzhou 350116, China}
    
	\author{Ke-Xiong Yan}
    \thanks{These authors contributed equally to this work.}
	\affiliation{Department of Physics, Fuzhou University, Fuzhou 350116, China}
	\affiliation{Fujian Key Laboratory of Quantum Information and Quantum Optics, Fuzhou University, Fuzhou 350116, China}

	\author{Yuan Qiu}
    \thanks{These authors contributed equally to this work.}
	\affiliation{Department of Physics, Fuzhou University, Fuzhou 350116, China}
	\affiliation{Fujian Key Laboratory of Quantum Information and Quantum Optics, Fuzhou University, Fuzhou 350116, China}

	\author{Yi-Tong Shi}
	\affiliation{Department of Physics, Fuzhou University, Fuzhou 350116, China}
	
	\author{Yan Xia}
	\affiliation{Department of Physics, Fuzhou University, Fuzhou 350116, China}
	\affiliation{Fujian Key Laboratory of Quantum Information and Quantum Optics, Fuzhou University, Fuzhou 350116, China}

	\author{Ye-Hong Chen}\thanks{yehong.chen@fzu.edu.cn}
	\affiliation{Department of Physics, Fuzhou University, Fuzhou 350116, China}
	\affiliation{Fujian Key Laboratory of Quantum Information and Quantum Optics, Fuzhou University, Fuzhou 350116, China}
	\affiliation{Quantum Information Physics Theory Research Team, Center for Quantum Computing, RIKEN, Wako-shi, Saitama 351-0198, Japan}
	\affiliation{Institute of Quantum Science and Technology, Yanbian University, Yanji 133002, China}

	\date{\today}
	
	\begin{abstract}
We propose an efficient and robust protocol for the generation of entanglement between a superconducting qubit and a squeezed cavity. By applying a parametric drive to the cavity coupled to the qubit, the dynamical evolution of the system is precisely described by an anisotropic Rabi model within a squeezed reference frame. Utilizing high-order time-averaging methods, we analytically derive the resonance conditions and the effective Rabi frequency for the high-order three-photon process. By implementing an adiabatic passage, slowly tuning the cavity frequency across the resonance, the system is steered into a maximally entangled state, e.g., between the three-photon state $\ket{g,3}$ and the qubit excited state $\ket{e,0}$ in the squeezed picture. Numerical simulation results confirm the high fidelity and robustness of the proposed protocol. Our method provides a practical pathway for generating complex non-Gaussian entangled states, which are of significant value for fault-tolerant quantum computation and quantum metrology beyond the standard quantum limit.
\end{abstract}

\maketitle

\section{Introduction}

Entanglement between discrete-variable and continuous-variable systems, known as hybrid entanglement, has emerged as a fundamental resource for hybrid quantum information processing~\cite{Jeong2014,Andersen2015,Huang2019,CampagneIbarcq2020,Ma2021,Sivak2023,PhysRevA.110.062402,Chang2025}. Unlike conventional entanglement, which pairs identical encodings such as two discrete qubits or two continuous optical beams, hybrid entanglement bridges the gap between different quantum degrees of freedom, effectively combining the precise logical addressability of discrete units with the vast Hilbert space of continuous modes~\cite{Ma2021,RevModPhys.93.025005}. This heterogeneous coupling bypasses the intrinsic non-Gaussianity limitations of pure continuous-variable systems~\cite{PhysRevLett.82.1784}, allowing the discrete component to drive the continuous mode into highly non-classical states~\cite{Hofheinz2009,McKay2015,Grimm2020}. As a result, hybrid entanglement has become a critical cornerstone for diverse applications, ranging from hardware-efficient bosonic quantum error correction~\cite{Gottesman2001,Terhal2020} to advanced quantum sensors that surpass the standard quantum limit~\cite{Giovannetti2011,Kristen2020}.

The performance of these applications is intrinsically tied to the specific quantum states involved in the hybrid entanglement. To date, research on the synthesis of hybrid entanglement has primarily concentrated on the coupling of qubits with either Gaussian states (such as coherent or squeezed vacuum states)~\cite{Vlastakis2013,PhysRevLett.120.093601} or ordinary Fock states~\cite{Eickbusch2022}. Entanglement between qubits and coherent states is a pivotal resource for fault-tolerant quantum computing, owing to the intrinsic robustness of coherent states against photon-loss errors~\cite{Lescanne2020}. Meanwhile, hybrid entanglement involving squeezed vacuum states is extensively utilized for detecting extremely weak forces or electromagnetic signals, leveraging the sub-vacuum noise fluctuations in a specific quadrature~\cite{PhysRevD.23.1693,Giovannetti2011,Schnabel2017}. Furthermore, the entanglement between qubits and Fock states is indispensable for implementing universal quantum logic gates due to its pronounced non-Gaussianity~\cite{PhysRevLett.106.060401,PhysRevA.92.040303}. In contrast, research on hybrid entanglement between the qubit and the squeezed multi-photon Fock state remains scarce. This specific class of hybrid entanglement reconciles the phase sensitivity of squeezed states~\cite{PhysRevD.32.400} with the non-Gaussianity of Fock states~\cite{Eickbusch2022}, offering significant potential for both fault-tolerant quantum information processing~\cite{Gottesman2001,Jia2025} and high-precision quantum metrology~\cite{Kristen2020,PhysRevResearch.6.033292}.

Despite its potential, the deterministic generation of entanglement between a qubit and a squeezed multi-photon Fock state remains a formidable challenge. In standard light-matter interactions governed by the Jaynes-Cummings (JC) model, the Rotating-Wave Approximation (RWA) constrains the system to single-excitation exchange processes, preserving the total excitation number~\cite{Shore1993,PhysRevLett.133.033603}.  Breaking this symmetry to facilitate high-order multi-photon transitions typically requires reaching the ultra-strong coupling regime, where counter-rotating terms become non-negligible, a condition often difficult to achieve in diverse experimental platforms~\cite{PhysRevA.110.043711}. To circumvent this limitation, parametric driving offers a powerful alternative by engineering synthetic non-linearities~\cite{PhysRevA.110.043711,Yan2025}. By modulating the system parameters, one can effectively introduce non-excitation-conserving terms that mimic the dynamics of an anisotropic Rabi model. This approach not only provides the necessary nonlinearity for multi-photon resonance but also intrinsically incorporates the squeezing degree of freedom. Consequently, the synergy between parametric modulation and high-order nonlinear coupling provides a robust pathway to synthesize hybrid entanglement involving complex non-Gaussian states, such as the squeezed three-photon Fock state. 

In this manuscript, we propose a protocol to generate entanglement between a qubit and a squeezed Fock state by applying a parametric drive to the JC model. The system dynamics in the rotating frame can be precisely described by an anisotropic Rabi model that incorporates counter-rotating terms. We theoretically derive the resonance conditions and the effective Rabi frequency for the three-photon Rabi oscillation in the squeezed reference frame. Subsequently, we present an adiabatic evolution protocol to achieve entanglement between the qubit and the squeezed cavity in the rotating frame by slowly sweeping the cavity frequency across the resonance point. Crucially, this evolution process corresponds to the deterministic generation of entanglement in the laboratory frame.

The rest of this paper is organized as follows. In Sec.~\ref{sec2}, we describe the system model and perform a squeezed reference frame transformation to derive an effective anisotropic Rabi Hamiltonian. Then, we analytically and numerically determine the three-photon resonance conditions and the resulting energy level splitting. Section~\ref{sec4} is dedicated to analyzing the critical role of the squeezing parameter $r$ in modulating the oscillation period and enhancing the state fidelity. In Sec.~\ref{sec5}, we present the adiabatic evolution protocol for the deterministic generation of hybrid entanglement, followed by a detailed assessment of the influence of environmental noise in Sec.~\ref{sec6}. Furthermore, a feasible experimental realization of our scheme using superconducting circuit QED platforms is discussed in Sec.~\ref{sec7}. Finally, we provide a summary and concluding remarks in Sec.~\ref{sec8}.

\section{Squeezed Frame Transformation and effective Hamiltonian} \label{sec2}

We consider a hybrid system consisting of a superconducting qubit coupled to a cavity subject to a parametric driving~\cite{PhysRevLett.120.093602}. In the frame rotating at half the drive frequency $\omega_p/2$, the Hamiltonian (setting $\hbar = 1$) is given by:
\begin{equation}
H = \delta_c a^\dagger a + \frac{\delta_q}{2} \sigma_z - \frac{\lambda}{2}(a^{\dagger 2} + a^2) + g(a^\dagger \sigma_- + a \sigma_+), \label{eq1}
\end{equation}
where $a$ ($a^\dagger$) is the cavity annihilation (creation) operator, $\sigma_z = |e\rangle\langle e| - |g\rangle\langle g|$ is the Pauli matrix, and $\sigma_\pm$ are the raising and lowering operators for the qubit, respectively. The parameters $\delta_c$ and $\delta_q$ respectively represent the detunings of the cavity and the qubit from the drive frequency $\omega_p$, while $\lambda$ denotes the parametric drive amplitude and $g$ is the bare coupling strength. To ensure system stability, we maintain the regime $|\delta_c| > \lambda$.

To simplify the dynamics of the system, we transform the Hamiltonian into a squeezed reference frame using the unitary operator $U_s(r) = \exp[\frac{r}{2}(a^2 - a^{\dagger 2})]$. The squeezing parameter $r$ is chosen to satisfy $\tanh 2r = \lambda/\delta_c$. In this representation, the transformed Hamiltonian is defined as $H_{aR} = U_s^\dagger H U_s - i U_s^\dagger \partial_t U_s$. As the squeezing parameter $r$ is time-independent, the dynamic correction term $U_s^\dagger \partial_t U_s$ vanishes, and the system is governed by an anisotropic Rabi model:
\begin{align}
H_{aR} =& \omega_c a^\dagger a + \frac{\omega_q}{2} \sigma_z \cr
&+ \lambda_1(a^\dagger \sigma_+ + a \sigma_-) + \lambda_2(a \sigma_+ + a^\dagger \sigma_-),
\label{eq5}
\end{align}
where the effective cavity frequency $\omega_c = \delta_c \text{sech}\,2r$ and the modified coupling strengths $\lambda_1 = g \sinh r$ and $\lambda_2 = g \cosh r$ are now explicitly dependent on the squeezing parameter.

This transformation provides a two-fold physical advantage. First, the effective frequency $\omega_c$ decreases with $r$, while the coupling strengths $\lambda_{1,2}$ are exponentially enhanced by the squeezing factor ($\sim e^r$) in the strong-squeezing limit ($e^r \gg 1$), effectively boosting the light-matter interaction. Second, the resulting Hamiltonian incorporates robust counter-rotating terms, specifically those proportional to $\lambda_2$, which enable the high-order multi-photon processes that would otherwise be forbidden under the standard rotating-wave approximation. Notably, the cavity vacuum state $|0\rangle$ in this squeezed frame corresponds to a squeezed vacuum state $S(r)|0\rangle$ in the original laboratory frame, ensuring that any state preparation within this frame naturally yields squeezed-state entanglement upon back-transformation.

To synthesize hybrid entanglement between a qubit and a squeezed three-photon Fock state, it is essential to realize a three-photon Rabi oscillation within the squeezed reference frame. Utilizing a high-order time-averaging method~\cite{PhysRevA.95.032124}, we derive an effective Hamiltonian to describe this high-order process and determine the corresponding Rabi frequency and the precise resonance condition. In the interaction picture, the anisotropic Rabi Hamiltonian $H_{aR}$ in Eq.~(\ref{eq5}) can be partitioned as:
\begin{equation}
H_I(t) = (h_1 e^{i\delta t} + h_2 e^{i\Delta t}) + \text{H.c.},
\end{equation}
where the operators are defined as $h_1 = \lambda_2 a \sigma_+$ and $h_2 = \lambda_1 a^\dagger \sigma_+$, with the corresponding detunings $\Delta = \omega_q + \omega_c$ and $\delta = \omega_q - \omega_c$. Under the three-photon resonance condition $\omega_q \approx 3 \omega_c$ (implying $\Delta \approx 2\delta$), and assuming the dispersive regime $|\lambda_n| \ll \delta$, we derive the second-order effective Hamiltonian~\cite{PhysRevA.110.043711}:
\begin{equation}
H_{\text{eff}}^{(2)} = \frac{\lambda_2^2}{2\omega_c} (a^\dagger a \sigma_z + \sigma_+ \sigma_-) + \frac{\lambda_1^2}{4\omega_c} (a^\dagger a \sigma_z - \sigma_- \sigma_+),
\end{equation}
and the third-order contribution:
\begin{equation}
H_{\text{eff}}^{(3)} = -\frac{\lambda_1 \lambda_2^2}{4\omega_c^2} (a^{\dagger 3} \sigma_- + a^3 \sigma_+).
\end{equation}
Combining these terms, the total effective Hamiltonian in the squeezed frame is given by:
\begin{equation}
H_{\text{eff}} = \omega_c a^\dagger a + \frac{\omega_q}{2} \sigma_z + H_{\text{eff}}^{(2)} + H_{\text{eff}}^{(3)}.
\label{eq6}
\end{equation}
The term $H_{\text{eff}}^{(3)}$ induces a direct coherent transition between the states $|e,n\rangle$ and $|g,n+3\rangle$ ($n=0,1,2,\ldots$), which, in the laboratory frame, corresponds to the transition dynamics between the qubit excited state and the squeezed multi-photon Fock state. The associated effective Rabi frequency $\Omega_{\text{eff}}$ is determined by the off-diagonal matrix element in the Hamiltonian.

\begin{figure}
	\centering
	\includegraphics[scale=0.5]{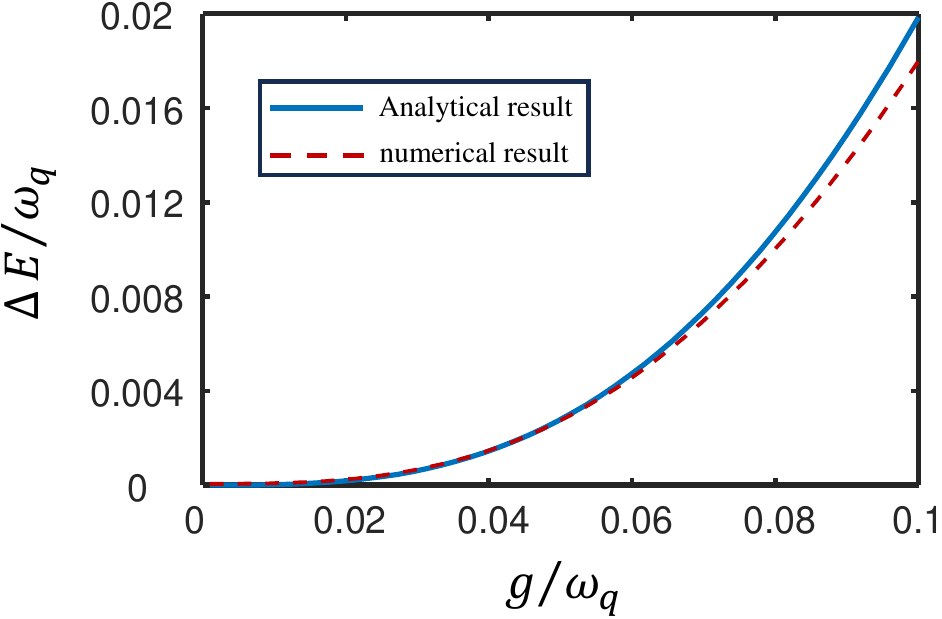}
	\caption{Comparison of the energy level splitting $\Delta E$ as a function of the interaction strength $g$ obtained from the analytical result (blue solid curve) and the numerical calculation (red dashed curve). The relevant parameters are $r = 0.9$, $\omega_c=\omega_c'$,  and $\omega_q = 1$.}
	\label{F1}
\end{figure}

For instance, in the subspace spanned by $\{|e,0\rangle, |g,3\rangle\}$, the effective Hamiltonian $H_{\text{eff}}$ in Eq.~(\ref{eq6}) can be rewritten as the matrix form:
\begin{equation}
H_{\text{sub}} = \begin{pmatrix} 
\frac{\omega_q}{2} + \frac{\lambda_2^2}{2\omega_c} & \ \ \ \ \ \ \ \  -\frac{\sqrt{6} \lambda_1 \lambda_2^2}{4 \omega_c^2} \\ 
-\frac{\sqrt{6} \lambda_1 \lambda_2^2}{4 \omega_c^2} &\ \ \ \ \ \ \ \  3\omega_c - \frac{\omega_q}{2} - \frac{3\lambda_2^2}{2\omega_c} - \frac{\lambda_1^2}{\omega_c} 
\label{eq8}
\end{pmatrix}.
\end{equation}
The associated effective Rabi frequency $\Omega_{\text{eff}}$ in this case reads
\begin{equation}
	\Omega_{\text{eff}} = \langle e,0|H_{\text{eff}}|g,3\rangle = -\frac{\sqrt{6} \lambda_1 \lambda_2^2}{4 \omega_c^2}.
\end{equation}
By equating the diagonal elements to satisfy the resonance condition, we obtain the shifted effective cavity frequency $\omega_c'$:
\begin{equation}
\frac{\omega_c'}{\omega_q} = \frac{1}{3} + (2\cosh^2 r + \sinh^2 r) \left(\frac{g}{\omega_q}\right)^2 + \mathcal{O}\left(\frac{g}{\omega_q}\right)^4.
\end{equation}
According to Eq.~(\ref{eq8}), the level splitting at the avoided crossing is given by $\Delta E = 2|\Omega_{\text{eff}}|$. This analytical result is subsequently validated by comparing it with the energy gap obtained through numerical diagonalization of the full Hamiltonian $H_{aR}$. The comparison is presented in Fig.~$\ref{F1}$ , which shows the analytical and numerical results for $\Delta E$ as a function of the fundamental coupling parameter $g$. The percentage difference between the two is less than $3\%$ for $g < 0.047$.

\begin{figure}
	\centering
	\includegraphics[scale=0.5]{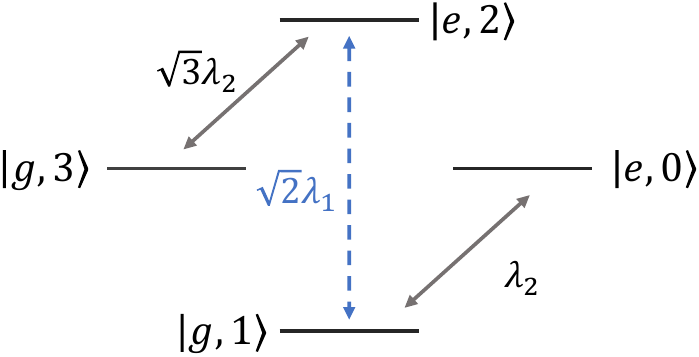}
	\caption{Schematic diagram illustrating the transition processes between the bare states $|e,0\rangle$ and $|g,3\rangle$ in squeezed picture. The excitation-number non-conserving processes are represented by blue dashed arrows, with $\lambda_2$, $\sqrt{2}\lambda_1$, and $\sqrt{3}\lambda_2$ denoting the corresponding transition matrix elements.}
	\label{F2}
\end{figure}

The effective Hamiltonian derived above demonstrates that a three-photon resonance can exist in the anisotropic Rabi model~\cite{PhysRevA.110.043711}. Figure~$\ref{F2}$ illustrates the transition between the states $|e,0\rangle$ and $|g,3\rangle$ in the squeezed picture, induced by an excitation-number non-conserving process. We further investigate this phenomenon through numerical simulations.

\begin{figure}
	\centering
	\includegraphics[scale=0.55]{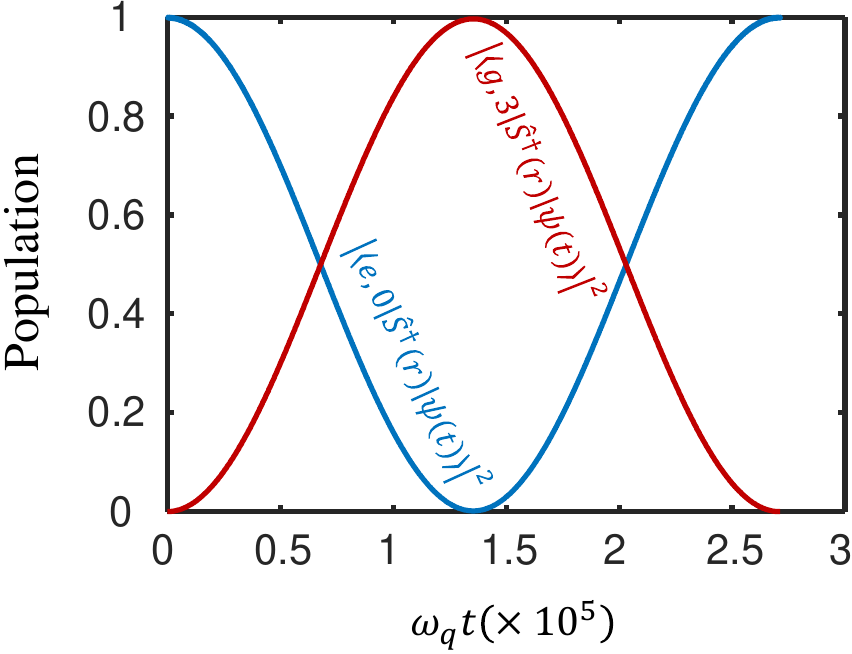}
	\caption{Time evolution of the probabilities $|\langle e,0 | \hat{S}^{\dagger}(r)\varphi(t) \rangle|^2$ (blue line) and  $|\langle g,3 |\hat{S}^{\dagger}(r) \varphi(t) \rangle|^2$ (red line), obtained from the numerical simulation of the dynamics governed by the Hamiltonian $H_{Ra}$ in Eq.~$(\ref{eq5})$. The highest probability of $\hat{S}(r)|g,3\rangle$ is $99.63\%$. The parameters are $r = 0.9$, $g = 0.01$, $\omega_c=\omega_c'$, and $\omega_q = 1$.}
	\label{F3}
\end{figure}

The simulation results demonstrate that under the given parameters $r = 0.9$, $g = 0.01$, the system exhibits perfect three-photon Rabi oscillations in the squeezed picture, as shown in Fig.~$\ref{F3}$. One can observe that the population of the squeezed three-photon state $\hat{S}(r)|g,3\rangle$ reaches a maximum of $99.63\%$, which verifies the effectiveness of our protocol. Further numerical simulations reveal that distinct Rabi oscillations persist for $r$ values spanning from $0$ to $2$, demonstrating that the oscillatory behavior remains robust over a wide parameter range.

\section{The Influence of the squeezing parameter r } \label{sec4}

\subsection{Effect on Oscillation Period}

Within this dynamical framework, the squeezing parameter $r$ serves as the primary factor governing the effective coupling strength $\Omega_{\text{eff}}$. To explicitly analyze its influence, the total Hamiltonian of the anisotropic Rabi model can be decomposed as $H = H_{\text{Rabi}} + H_{\text{ARabi}}$~\cite{PhysRevLett.120.093602}, where:
\begin{align}
H_{\text{Rabi}} &= \omega_c  a^\dagger a + \frac{\delta_q}{2} {\sigma}_z + \frac{g}{2} e^{r} ({a}^\dagger + {a})({\sigma}_+ + {\sigma}_-), \label{eq:HRabi} \\
{H}_{\text{ARabi}} &= -\frac{g}{2} e^{-r} ({a}^\dagger - {a})({\sigma}_+ - {\sigma}_-). \label{eq:HErr}
\end{align}
Equations \eqref{eq:HRabi} and \eqref{eq:HErr} reveal the dual role of the parameter $r$ in modulating the properties of the system. In the Hamiltonian $\hat{H}_{\text{Rabi}}$, the light-matter coupling is exponentially enhanced by the factor $e^{r}$, which significantly boosts the effective Rabi frequency $\Omega_{\rm eff}$ of the three-photon resonance in the squeezed picture. For a fixed fundamental coupling $g$, an increase in $r$ leads to a notable reduction in the oscillation period $t_{f}$, thereby expediting the coherent evolution. 

\begin{figure}
	\centering
	\includegraphics[scale=0.55]{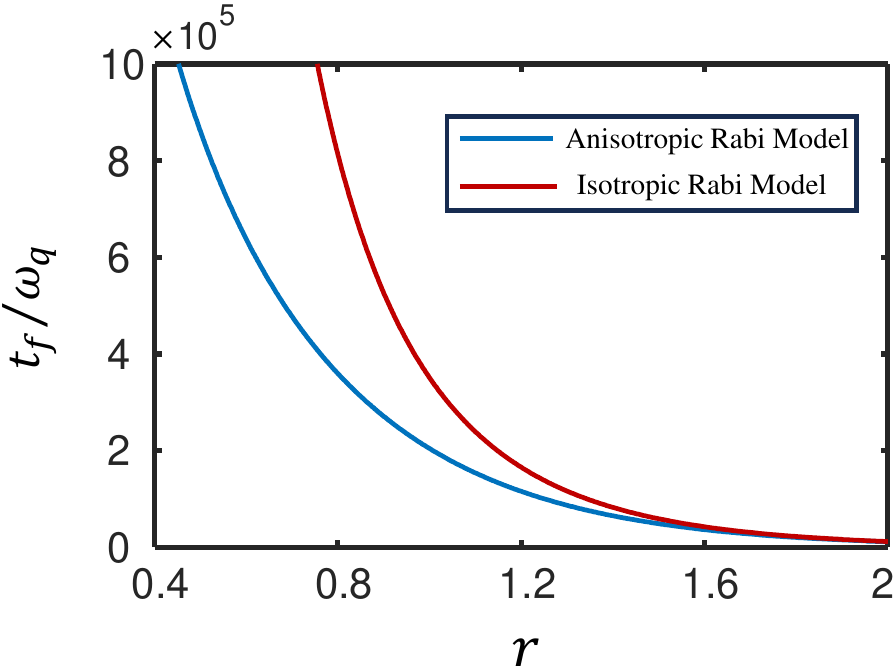}
	\caption{Oscillation period $t_f$ of the isotropic Rabi model (red line) and the anisotropic Rabi model (blue line) as a function of the squeezing parameter $r$. The fundamental coupling parameter is $g = 0.01\omega_q$, $\omega_c=\omega_c'$, and $\omega_q = 1$. For ease of observation, the range of $r$ is set to $[0.4, 2]$.}
	\label{F4}
\end{figure}

Simultaneously, the anisotropic component represented by $\hat{H}_{\text{ARabi}}$ is exponentially suppressed by the factor $e^{-r}$. Consequently, as $r$ increases, the contribution of the anisotropic term becomes negligible, driving a transition of the system from an anisotropic Rabi regime toward an isotropic one. These features and their associated oscillatory signatures are illustrated in Fig.~\ref{F4}, where the oscillation period curves of the two models eventually coincide once $r$ exceeds a certain threshold. Thus, the parameter $r$ acts as a versatile control knob, enabling both the fine-tuning of the oscillation frequency and the strategic manipulation of the symmetry class of the model.

\subsection{Effect on the Fidelity of the Three-Photon State}

\begin{figure}
	\centering
	\includegraphics[scale=0.55]{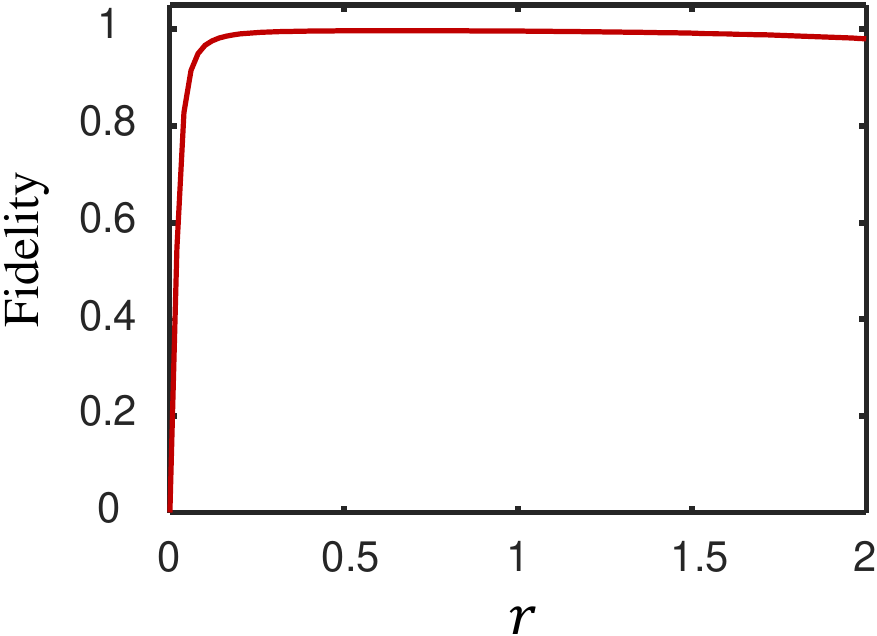}
	\caption{Fidelity of the three-photon state $|g,3\rangle$ as a function of parameter $r$, with $g = 0.01\omega_q$, $\omega_c=\omega_c'$, and $\omega_q = 1$.}
	\label{F5}
\end{figure}

In this section, with the fundamental coupling strength $g$ fixed, the value of parameter $r$ has a significant impact on the fidelity of the target three-photon state $\ket{g,3}$. To systematically investigate the variation of fidelity with parameter $r$, the maximum fidelity of the $\ket{g,3}$ state as a function of $r$ is calculated and plotted, as shown in Fig.~$\ref{F5}$. The figure clearly shows that as parameter $r$ approaches $0.65$, the fidelity peaks at a value close to $1$, indicating that the target three-photon state can be prepared with high precision under this parameter condition.

This moderate requirement for $r$ is of significant practical importance. In state-of-the-art circuit QED experiments, achieving and maintaining extremely high squeezing levels remains a formidable challenge due to higher-order nonlinearities, pump depletion, and increased susceptibility to decoherence in strong driving regimes \cite{PhysRevLett.107.113601,f2y1-scgw}. While record-breaking squeezing has been reported in specialized setups, typical operational squeezing levels in qubit-coupled systems are often limited to moderate values to preserve system stability \cite{PhysRevX.7.041011}. Our results demonstrate that a squeezing level of $r \approx 0.65$ (corresponding to approximately $5.6$~dB of squeezing) is sufficient to synthesize the target three-photon state with high precision. Since this value falls well within the accessible range of current Josephson parametric devices, our protocol offers a highly feasible and robust pathway for experimental implementation under realistic conditions.

\begin{figure*}[t]
	\centering
	\includegraphics[scale=0.8]{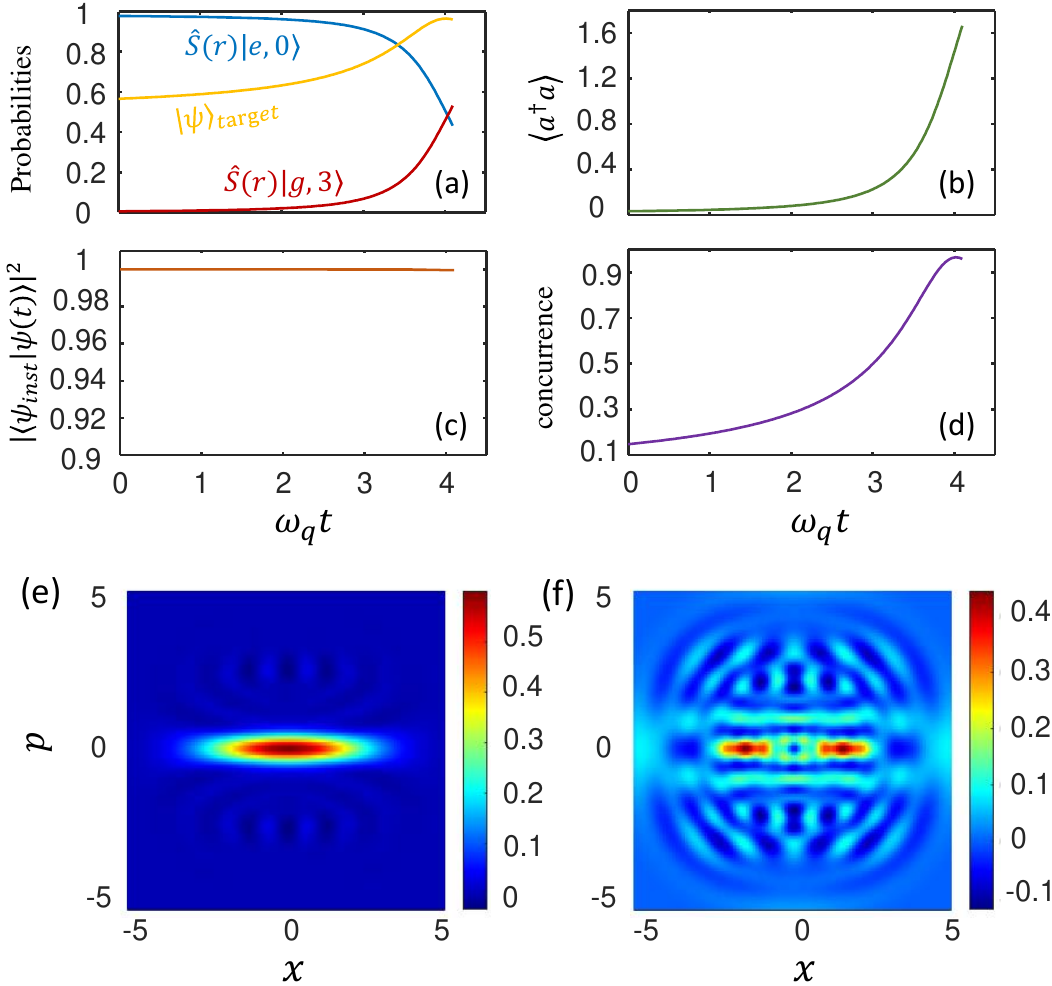}
	\caption{Adiabatic Evolution Dynamics. (a) Time evolution of the populations for the states $\hat{S}(r)|e,0\rangle$ (blue solid line),  $\hat{S}(r)|g,3\rangle$ (red dashed line), and the target state $\ket{\psi_{\rm target}}=|\Psi\rangle_{\text{Lab}}$ in the lab frame. (b) Expectation value of the photon number $\langle a^\dagger a \rangle$ in the squeezed picture. (c) Adiabaticity check via the projection probability $|\langle \psi_{\text{inst}} | \psi(t) \rangle|^2$ for the target adiabatic branch, where $\ket{\psi_{\text{inst}}}$ is the fourth excited eigenstate of the Hamiltonian $H(t)$. (d)  The concurrence of  $\ket{\psi(t)}$ as a function of evolution time. (e) and (f) are the Wigner phase-space distribution of the initial state and the final state in the lab frame. All results are plotted as functions of $\omega_q t$ with $\omega_q=1$, $g = 0.06\omega_q$, $r=0.9$, $v = 0.05\Omega_{\rm eff}^2$, and $\omega_c(0) = \omega_c'-0.01\omega_q$.}
	\label{F7}
\end{figure*}

\section{Adiabatic Preparation of Hybrid Entanglement} \label{sec5}

Based on the avoided level crossing analyzed in the previous sections, we propose an adiabatic protocol to deterministically generate qubit-resonator entanglement. The protocol relies on a time-dependent modulation of the effective cavity frequency, $\omega_c(t) = \omega_{c}(0) + vt$, where $v$ is the sweep rate. In the squeezed reference frame, the system dynamics are governed by the time-dependent anisotropic Rabi Hamiltonian~\cite{PhysRevA.92.023842}:
\begin{equation}
H(t) = \omega_c(t) a^\dagger a + \frac{\omega_q}{2} \sigma_z + \lambda_1(a^\dagger \sigma_+ + a \sigma_-) + \lambda_2(a \sigma_+ + a^\dagger \sigma_-),
\label{eq_H_adiabatic}
\end{equation}
where the coupling strengths $\lambda_1 = g \sinh r$ and $\lambda_2 = g \cosh r$ are determined by the squeezing parameter $r$. The system is initialized in the instantaneous eigenstate that corresponds to the bare state $|e, 0\rangle$ at a large positive detuning, such that $\Delta(0) = \omega_c(0) - \omega_c' \gg |\Omega_{\text{eff}}|$.

As $\omega_c(t)$ is slowly tuned through the three-photon resonance point $\omega_c'$, the system follows its instantaneous eigenstate along the adiabatic branch. According to the Landau-Zener theorem~\cite{Zener1932-pi,noauthor_1965-na}, the probability of a successful adiabatic transition is governed by the adiabatic parameter $\eta = |\Omega_{\text{eff}}|^2 / v$. To ensure deterministic preparation, we require the adiabatic condition $\eta \gg 1$, implying that the sweep must be sufficiently slow relative to the square of the effective coupling strength. 

At the resonance point $\Delta(t) = 0$, the state of the system in the squeezed frame becomes a Bell-like entangled state:
\begin{equation}
|\Psi_{\text{res}}\rangle_{\text{SF}} = \frac{1}{\sqrt{2}} (|e, 0\rangle - |g, 3\rangle).
\end{equation}
By halting the sweep at this point, we obtain a stable entangled manifold. Upon transforming back to the laboratory frame via the inverse squeezing operator $S(r) = \exp[\frac{r}{2}(a^{\dagger 2} - a^2)]$, this state is mapped to:
\begin{equation}
|\Psi\rangle_{\text{Lab}} = \frac{1}{\sqrt{2}} \left( |e\rangle \otimes S(r)|0\rangle - |g\rangle \otimes S(r)|3\rangle \right).
\label{eq_lab_state}
\end{equation}
In the laboratory frame, Eq.~(\ref{eq_lab_state}) represents the hybrid entanglement between a discrete qubit and a non-Gaussian squeezed three-photon Fock state. 

Figure~\ref{F7} illustrates the numerical simulations of this adiabatic process. As shown in Fig.~\ref{F7}(a), the population is smoothly transferred from $\hat{S}(r)|e,0\rangle$ to $\hat{S}(r)|g,3\rangle$ in the squeezed frame, reaching an equal superposition at the resonance point.  The maximum entangled state $\ket{\psi_{\rm target}}$ reaches the highest population up to 1. Fig.~\ref{F7}(b) depicts the evolution of the average photon number $\langle a^\dagger a \rangle$, which rises toward $1.5$ at the final time, confirming the formation of the $\hat{S}(r)|g,3\rangle$ component. The adiabaticity is further verified in Fig.~\ref{F7}(c), where the fidelity between the evolved state and the instantaneous eigenstate remains near unity throughout the sweep. The time evolution of the entanglement during the adiabatic process is presented in Fig.~\ref{F7}(d). It is observed that the entanglement reaches its maximum at the end of the adiabatic passage, signaling the successful preparation of hybrid entanglement between the qubit and the squeezed three-photon state. Figures~\ref{F7}(e) and 6(f) display the Wigner phase-space distributions of the cavity field in the laboratory frame for the initial state and final state, respectively. The results clearly illustrate the transition of the cavity from an initial squeezed vacuum state to a squeezed three-photon Fock state. Notably, the squeezing parameter $r$ acts as the dominant control knob: increasing $r$ exponentially enhances $\Omega_{\text{eff}}$, thereby allowing for a faster sweep rate while maintaining high-fidelity state preparation.

\section{Influence of environmental noise} \label{sec6}

In realistic cQED implementations, the system is inevitably coupled to its environment. Since the bare coupling strength in our model remains within the JC regime ($g \ll \omega_c$), the dissipative dynamics can be accurately described by the standard Lindblad master equation~\cite{10.1093/acprof:oso/9780199213900.001.0001}:
\begin{equation}
\frac{d\rho(t)}{dt} = -i [H(t), \rho(t)] + \kappa \mathcal{D}[a] \rho + \gamma \mathcal{D}[\sigma_-] \rho,
\label{eq:master}
\end{equation}
where $\rho(t)$ represents the density matrix of the hybrid system. The term $H(t)$ is the time-dependent Hamiltonian driving the adiabatic evolution. The Lindblad superoperator $\mathcal{D}[L]\rho = L\rho L^\dagger - \frac{1}{2}\{L^\dagger L, \rho\}$ describes the Markovian decoherence processes. Specifically, $\kappa$ denotes the decay rate of the cavity mode, representing photon loss to the environment via the annihilation operator $a$. The parameter $\gamma$ characterizes the longitudinal relaxation rate of the superconducting qubit, mediated by the lowering operator $\sigma_- = |g\rangle \langle e|$.

To evaluate the experimental feasibility of our protocol, we perform numerical simulations of Eq.~(\ref{eq:master}) with a squeezing parameter $r = 0.9$ and realistic dissipation rates: $\kappa = 0.009\omega_q$ and $\gamma = 0.0001\omega_q$. We focus on two key metrics: the concurrence, which quantifies the degree of qubit-resonator entanglement, and the fidelity of the target state $\ket{\psi_{\rm target}}$, which measures the overlap between the dissipative state and the ideal maximally entangled state $|\Psi\rangle_{\text{Lab}} = \frac{1}{\sqrt{2}} \left( |e\rangle \otimes S(r)|0\rangle - |g\rangle \otimes S(r)|3\rangle \right)$.

The numerical results demonstrate that  despite the presence of cavity loss and  qubit dissipation, the fidelity and the concurrence can reach 0.77 and 0.71, respectively. This confirms the experimental feasibility of our scheme under current circuit QED parameters.

\begin{figure}
	\centering
	\includegraphics[scale=0.5]{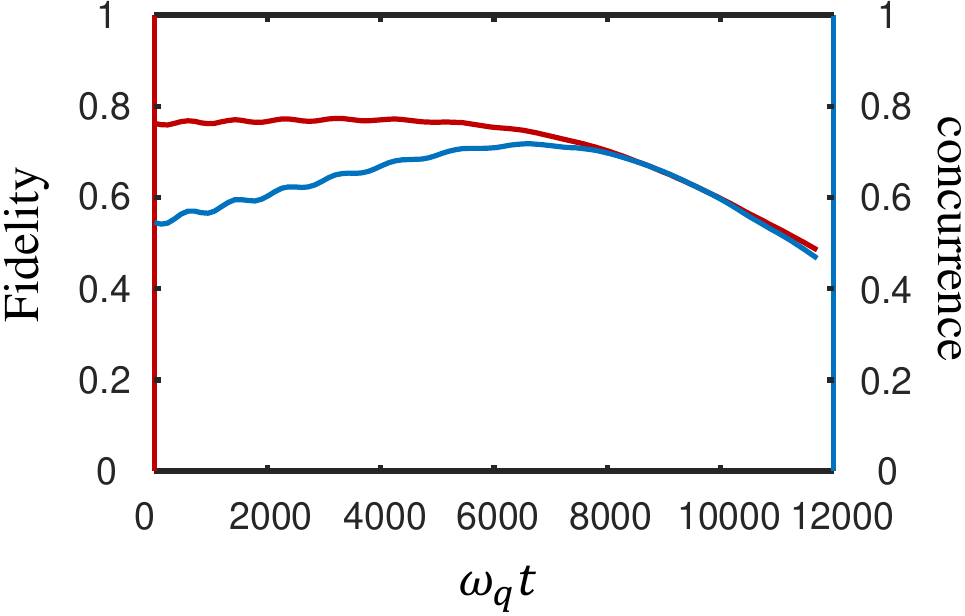}
	\caption{Time dependence of the fidelity and the concurrence. The relevant parameters are: $\kappa = 5\times10^{-5}\omega_q$, $\gamma = 3\times10^{-5}\omega_q$, $r = 0.5$, and $\omega_q=1$.}
	\label{F6}
\end{figure}

\section{Experimental implementation in superconducting circuits} \label{sec7}

\begin{figure}
	\centering
	\includegraphics[scale=0.6]{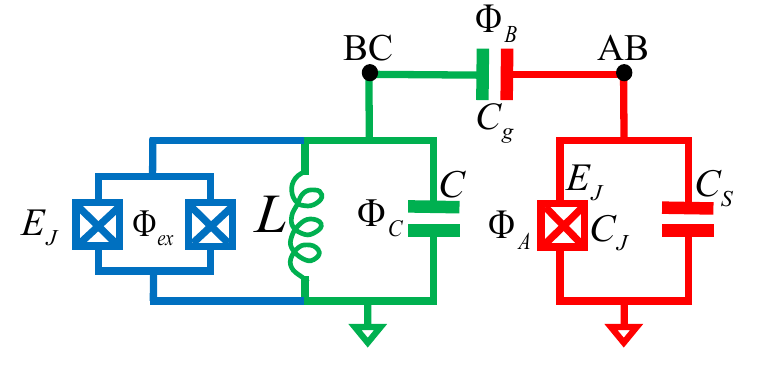}
	\caption{Possible implementation in superconducting circuits. Different colors represent distinct components: red denotes the superconducting qubit, green represents the resonator, and blue indicates the external parametric drive.}
	\label{F8}
\end{figure}

In this section, we give a possible experimental realization scheme in circuit QED platforms~\cite{Clerk2020,Qiu_2025}. The circuit depicted in Fig.~\ref{F8} comprises three functional modules: the parametric drive source (blue), the $LC$ resonator (green), and the superconducting Transmon qubit (red). To derive the effective Hamiltonian, we define the generalized coordinates as the node fluxes $\Phi_r$ at node BC and $\Phi_q$ at node AB. The corresponding voltages are given by $\dot{\Phi}_r$ and $\dot{\Phi}_q$, respectively.

The kinetic energy $\mathcal{T}$ of the system is stored in the capacitive network. By defining $C_{\Sigma r} = C + C_g$ and $C_{\Sigma q} = C_s + C_J + C_g$ as the total capacitances at the resonator and qubit nodes, the kinetic energy is expressed as:
\begin{equation}
\mathcal{T} = \frac{1}{2} C_{\Sigma r} \dot{\Phi}_r^2 + \frac{1}{2} C_{\Sigma q} \dot{\Phi}_q^2 - C_g \dot{\Phi}_r \dot{\Phi}_q.
\end{equation}
The cross-term $C_g \dot{\Phi}_r \dot{\Phi}_q$ represents the capacitive coupling between the two subsystems. The potential energy $V$ is determined by the inductive elements and the nonlinear Josephson junctions, given by:
\begin{eqnarray}
    V &=& \frac{\Phi_r^2}{2L} - 2E_J \cos\left[ \frac{\pi \Phi_{\text{ex}}(t)}{\Phi_0} \right] \cos\left( \frac{2\pi \Phi_r}{\Phi_0} \right)\\ \nonumber
    &&- E_J \cos\left( \frac{2\pi \Phi_q}{\Phi_0} \right),
\end{eqnarray}
where $\Phi_0 = h/2e$ is the magnetic flux quantum. The blue SQUID loop is biased by a time-dependent external flux $\Phi_{\text{ex}}(t)$, which serves as the source of the parametric drive.

By defining the conjugate charges as $Q_i = \partial \mathcal{L} / \partial \dot{\Phi}_i$, we perform a Legendre transformation to obtain the classical Hamiltonian. In the weak coupling limit ($C_g \ll C_{\Sigma r}, C_{\Sigma q}$), the Hamiltonian is approximately:
\begin{equation}
H = \frac{Q_r^2}{2C_{\Sigma r}} + \frac{Q_q^2}{2C_{\Sigma q}} + \frac{Q_r Q_q}{C_{cc}} + \frac{\Phi_r^2}{2L} + U_{\text{nl}}(\Phi_r, \Phi_q, t),
\end{equation}
where $U_{\text{nl}}$ incorporates the nonlinear junction potentials and $C_{cc}$ denotes the effective coupling capacitance.

We perform canonical quantization by introducing the bosonic annihilation (creation) operators $a$ ($a^\dagger$) for the resonator mode and the Pauli operators $\sigma_i$ for the qubit. The flux and charge operators for the resonator are defined as:
\begin{equation}
\hat{\Phi}_r = \Phi_{\text{zpf}}(a + a^\dagger), \quad \hat{Q}_r = -i Q_{\text{zpf}}(a - a^\dagger),
\end{equation}
where $\Phi_{\text{zpf}} = \sqrt{\hbar Z_c / 2}$ is the zero-point fluctuation of the flux and $Z_c = \sqrt{L/C_{\Sigma r}}$ is the characteristic impedance. For the parametric driving, we expand the SQUID potential $\cos(2\pi \Phi_r / \Phi_0)$ to the second order in $\hat{\Phi}_r$. By modulating the external flux $\Phi_{\text{ex}}(t)$ at approximately twice the resonator frequency ($\omega_p \approx 2\omega_c$), the nonlinear term $E_{J,\text{eff}}(t) \hat{\Phi}_r^2$ generates the parametric pump:
\begin{equation}
H_{p} = -\frac{\lambda}{2}(a^2 + a^{\dagger 2}).
\end{equation}
When $E_J/E_C \gg 1$, the qubit is restricted to its lowest two energy levels, described by:
\begin{equation}
H_q = \frac{1}{2} \delta_q \sigma_z.
\end{equation}
The capacitive interaction $Q_r Q_q$ translates to the dipole-like coupling:
\begin{equation}
H_{\text{int}} = g(a + a^\dagger)(\sigma_+ + \sigma_-).
\end{equation}

Finally, by setting $\hbar = 1$ and transforming into a reference frame rotating at half the pump frequency $\omega_p/2$, we arrive at the model Hamiltonian in Eq.~(\ref{eq1}):
\begin{equation}
H = \delta_c a^\dagger a + \frac{\delta_q}{2} \sigma_z - \frac{\lambda}{2} (a^2 + a^{\dagger 2}) + g(a + a^\dagger)(\sigma_+ + \sigma_-).
\end{equation}
The mapping between the physical circuit components and the model parameters is established such that the resonator detuning $\delta_c$ is derived from the $LC$ frequency relative to the drive frequency, the parametric amplitude $\lambda$ is determined by the flux modulation $\Phi_{\text{ex}}(t)$ in the blue SQUID loop, and the coupling strength $g$ is fixed by the mutual capacitance $C_g$.

\section{Conclusion} \label{sec8}

We have proposed a scheme for generating hybrid entanglement between a qubit state and a squeezed Fock state via three-photon resonance and adiabatic evolution. By applying a parametric drive to the cavity, the JC model is mapped to an anisotropic Rabi model in a squeezed reference frame. Using the high-order time-averaging method, we have analytically derive the resonance condition for three-photon Rabi oscillations and the corresponding effective coupling frequency. Our analysis shows that the target state can be efficiently prepared with a squeezing level of approximately 5.6 dB. Through numerical simulations, we confirm that the protocol achieves high fidelity and robustness under ideal conditions, while still maintaining appreciable entanglement in the presence of practical dissipative effects, such as cavity damping and qubit relaxation.This work provides a feasible approach for generating entangled resources that combine the phase-sensitivity of squeezed states with the non-Gaussian characteristics of Fock states in superconducting quantum circuits, offering significant potential for advancing fault-tolerant quantum computation and quantum precision measurement.

\section*{Acknowledgements}
Y.-T. S. was supported by the Student Research Training Program (SRTP) of China under Grant No.~202510386052.
Y.-H.C. was supported by the National Natural Science Foundation of China under Grant No.~12304390 and 12574386, the National Postdoctoral Overseas Talent Recruitment Program of China, the Fujian 100 Talents Program, and the Fujian Minjiang Scholar Program. Y. X. was supported by the National Natural Science
Foundation of China under Grant No.~62471143, the
Key Program of National Natural Science Foundation
of Fujian Province under Grant No.~2024J02008.

	\bibliography{reference}
\end{document}